\documentclass[conference]{IEEEtran}
\IEEEoverridecommandlockouts
\usepackage{cite}
\usepackage{amsmath,amssymb,amsfonts}
\usepackage{algorithmic}
\usepackage{graphicx}
\usepackage{textcomp}
\usepackage{xcolor}

\newtheorem{lemma}{Lemma}
\newtheorem{theorem}{Theorem}
\usepackage{bm}
\def\BibTeX{{\rm B\kern-.05em{\sc i\kern-.025em b}\kern-.08em
    T\kern-.1667em\lower.7ex\hbox{E}\kern-.125emX}}
\begin{document}

\title{Machine Learning-Based Secret Key Generation for  IRS-assisted  Multi-antenna Systems
}

\author{
\IEEEauthorblockN{
Chen~Chen\IEEEauthorrefmark{1},
Junqing~Zhang\IEEEauthorrefmark{1},
Tianyu~Lu\IEEEauthorrefmark{2},
Magnus~Sandell\IEEEauthorrefmark{3},
Liquan~Chen\IEEEauthorrefmark{2}
}

\IEEEauthorblockA{
\IEEEauthorrefmark{1}
Department of Electrical Engineering and Electronics, University of Liverpool, Liverpool, L69 3GJ, United Kingdom\\ 
Emails: c.chen77@liverpool.ac.uk, Junqing.Zhang@liverpool.ac.uk\\
\IEEEauthorrefmark{2}
School of Cyber Science and Engineering, Southeast University, Nanjing, 210096, China\\ 
Emails: effronlu@seu.edu.cn, lqchen@seu.edu.cn\\
\IEEEauthorrefmark{3}
Bristol Research and Innovation Laboratory, Toshiba Research Europe Ltd., Bristol, BS1 4ND, United Kingdom\\
Email: magnus.sandell@toshiba-bril.com}}


\maketitle

\begin{abstract}
Physical-layer key generation (PKG) based on wireless channels is a lightweight technique to establish secure keys  between legitimate communication nodes. Recently, intelligent reflecting surfaces (IRSs) have been leveraged to enhance the performance of PKG in terms of secret key rate (SKR), as it can reconfigure the wireless propagation  environment and introduce more channel randomness. In this paper, we investigate an IRS-assisted PKG system, taking into account the channel spatial correlation at both the base station (BS) and the IRS. Based on the considered system model, the closed-form expression of SKR is  derived analytically. Aiming to maximize the SKR, a joint design problem of the BS's precoding matrix and the IRS’s reflecting coefficient vector is formulated. To address this high-dimensional non-convex optimization problem, we propose a novel unsupervised deep neural network (DNN) based algorithm with a simple structure. Different from most previous works that adopt the iterative optimization to solve the problem, the proposed DNN based algorithm directly obtains the BS precoding and IRS phase shifts as the output of the DNN. Simulation results reveal that the proposed DNN-based algorithm outperforms the benchmark methods with regard to SKR.
\end{abstract}

\begin{IEEEkeywords}
Physical-layer key generation, intelligent reflecting surfaces, deep neural network
\end{IEEEkeywords}

\section{Introduction}
Along with the surge in data traffic envisioned by sixth generation (6G) communication, data security risks emerge due to the broadcast nature of wireless transmissions~\cite{Magzine1}.  In this context, physical-layer key generation (PKG) has been proposed as a promising technique that does not rely on conventional cryptographic  approaches and is information-theoretically secure~\cite{Access1}. PKG utilizes the reciprocity of wireless channels during the channel coherent time to generate symmetric keys from bidirectional channel probings. The spatial decorrelation and channel randomness ensure the security of the generated keys.

However, PKG faces serious challenges in harsh propagation environments. The transmission link between two legitimate nodes may experience non-line-of-sight (NLOS) propagation conditions. In this case,  the channel estimation will suffer from a low signal-to-noise ratio (SNR), resulting in a low  secret key rate (SKR).  Besides, in a quasi-static environment that lacks randomness, the achievable SKR is limited due to low channel entropy \cite{Access1}. 

To address these challenges, a new paradigm, called intelligent reflecting surface (IRS), offers attractive solutions~\cite{CL1, TVT1, TIFS2}.
IRS consists of low-cost passive reflecting elements that can dynamically adjust their amplitudes and/or phases to reconfigure the  wireless propagation environments~\cite{wu2021intelligent}.  The reflection channels provided by IRS are capable of enhancing the received signals at legitimate users and introducing rich channel randomness. The work in~\cite{CL1} improved the SKR of an IRS-aided single-antenna system by randomly configuring the IRS's phase shifts, in which closed-form expressions of the lower and upper bounds of the SKR were provided. In~\cite{TVT1}, the IRS reflecting coefficients were optimized to maximize the minimum achievable SKR. Successive convex approximation was adopted to address the non-convex optimization objective. In~\cite{TIFS2}, the authors considered a multi-user scenario and designed IRS phase shifts to maximize the sum SKR. Generally, the optimal configuration of IRS is a complex non-convex optimization problem and requires iterative optimization. This problem is exacerbated in multi-antenna systems that require a joint design of multi-antenna precoding and IRS phase shifting. 

Recently, machine learning has
emerged as one of the key techniques of 6G and has been employed to address mathematically intractable non-convex optimization problems~\cite{Survey1}. There have been several works~\cite{Access2, JSAC1} leveraging machine learning to optimize IRS reflecting beamforming with the aim of maximizing downlink transmission rates. A recent work~\cite{jiao2021machine} proposed a machine learning based adaptive quantization method to balance key generation rate and bit disagreement ratio in an IRS-aided single-antenna system. Up to now, the employment of machine learning to jointly optimize BS precoding and IRS reflecting coefficient vector in an IRS-aided PKG system has not been investigated yet.

To the best of our knowledge, this is the first attempt to exploit machine learning in PKG for an IRS-assisted multi-antenna system. Our
major contributions are summarized as follows: 
\begin{itemize}
	\item We propose a new PKG framework for an IRS-assisted multi-antenna system and derive the closed-form expression of SKR.
	\item The optimization problem of SKR maximization is formulated and a water-filling algorithm based baseline solution is developed.
	\item We novelly propose to obtain the optimal configuration of BS precoding and IRS phase shifting by using unsupervised deep neural networks (DNNs), referred to as ``PKG-Net''. Simulations demonstrate that the proposed PKG-Net can achieve a higher SKR than other benchmark methods. 
\end{itemize}

\emph{Notations:} In this paper, scalars are represented by italic letters, vectors
by boldface lower-case letters,
and matrices  by boldface upper-case letters. $\mathbf{V}^{T}$, $\mathbf{V}^{H}$ and $\mathbf{V}^{*}$ are the transpose, conjugate transpose
and conjugate of a matrix $\mathbf{V}$, respectively.
$\mathbb{E}\{\cdot\}$ is
the statistical expectation. $\mod(\cdot)$ and $\lfloor\cdot\rfloor$ denote modulus operator and  floor function, respectively. $[\mathbf{V}]_{i,j}$ denotes the $(i,j)$-th element of a matrix $\mathbf{V}$. $\mathcal{CN}(\mu,\sigma^2)$ represents a circularly symmetric complex Gaussian  distribution with mean $\mu$ and variance $\sigma^2$.  $\text{diag}(\mathbf{v})$ is a diagonal matrix with the entries of $\mathbf{v}$ on its main diagonal and $\text{vec}(\mathbf{V})$ is the vectorization of a matrix $\mathbf{V}$. $||\cdot||_F$ denotes the Frobenius norm. $\odot$ and $\otimes$ are the Hadamard product and Kronecker product, respectively. 

\section{System Model}
\subsection{System Overview}
\begin{figure}[!t]
\centerline{\includegraphics[width=3.4in]{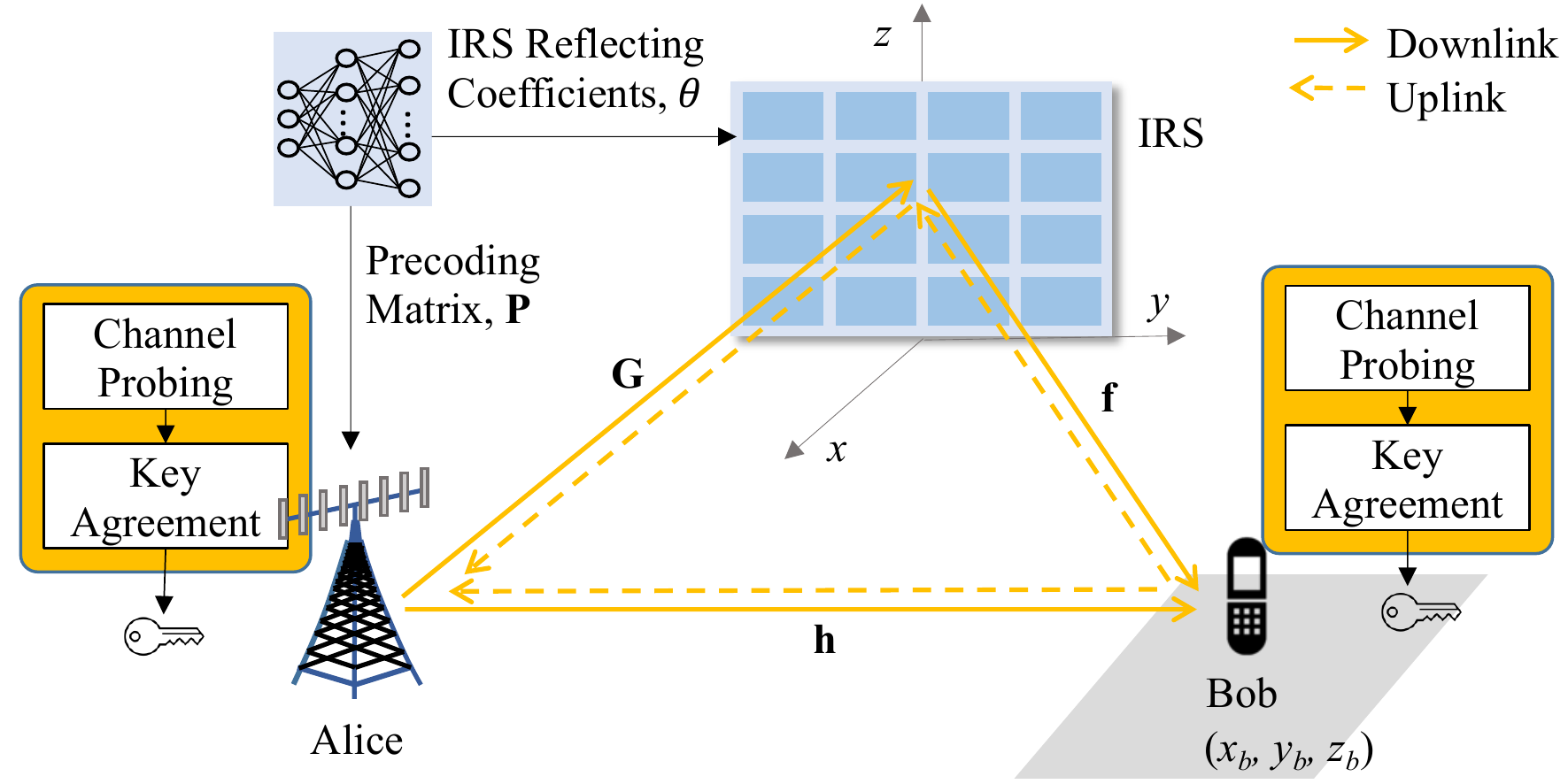}}
\caption{System model.}
\label{fig:channel}
\end{figure}

We consider a narrowband scenario as shown in Fig. \ref{fig:channel}, which contains a multi-antenna BS, Alice, a single-antenna user equipment (UE), Bob, and an IRS. We assume that the BS is a uniform linear array with $M$ antennas, and the IRS is a uniform planar array, which consists of $L=L_{\mathrm{H}}\times L_{\mathrm{V}}$ passive reflecting elements with $L_{\mathrm{H}}$ elements per row and $L_{\mathrm{V}}$ elements per column. To secure the communication, Alice and Bob perform PKG with the assistance of the IRS.

As shown in Fig. \ref{fig:channel}, the PKG protocol consists of two phases: channel probing and key agreement. During the channel probing phase, Alice and Bob perform bidirectional channel measurements by transmitting pilot signals to each other.   We assume a time-division duplexing (TDD) protocol, and thus Alice and Bob can observe reciprocal channel information. During the key agreement phase, Alice and Bob quantize the channel observations into binary key bits. Then the disagreed key bits are corrected by information reconciliation and the leaked information is eliminated by privacy amplification~\cite{Access1}. In this paper, we focus on the channel probing phase.


\subsection{Channel Model}
We denote the BS-IRS, IRS-UE and BS-UE channels by $\mathbf{G}\in\mathbb{C}^{M\times L}$, $\mathbf{f}\in\mathbb{C}^{L\times 1}$ and $\mathbf{h}\in\mathbb{C}^{M\times 1}$, respectively. In practice, there exist spatial correlations among BS antennas and IRS elements. For simplicity, we adopt the spatial correlation model in isotropic scattering environments \cite{WCL1}. The ($n,m$)-th element of the spatial correlation matrix at the IRS is given by~\cite{WCL1}
\begin{align}
[\mathbf{R}_{\mathrm{I}}]_{n,m}=\frac{\sin(\frac{2\pi}{\lambda}||\mathbf{u}_n-\mathbf{u}_m||_2)}{\frac{2\pi}{\lambda}||\mathbf{u}_n-\mathbf{u}_m||_2},~n,m=1,\dots,L,
\end{align} 
where $\lambda$ is the wavelength, and $\mathbf{u}_n$ denotes the location of the $n$-th element and is computed as
$\mathbf{u}_n = \left[0, y_{n}\Delta, z_{n}\Delta\right]^{T}$, where $y_{n}=\mod(n-1, L_{\mathrm{H}})$ and $z_{n}=\left \lfloor (n-1)/L_{\mathrm{H}} \right \rfloor$ are the horizontal and vertical indices of the $n$-th element, respectively, and $\Delta$ is the element spacing. The spatial correlation matrix at the BS is represented by $\mathbf{R}_{\mathrm{B}}$ with the ($n,m$)-th element given by 
\begin{align}
	[\mathbf{R}_{\mathrm{B}}]_{m,n}=\eta^{|m-n|},
\end{align}
 where $\eta$ is the correlation coefficient among BS antennas~\cite{GLOBECOM1}. Then the channels can be described as 
\begin{align}
\mathbf{G}&=\mathbf{R}_{\mathrm{B}}^{1/2}\widetilde{\mathbf{G}}\mathbf{R}_{\mathrm{I}}^{1/2}, \\
\mathbf{f}&=\mathbf{R}_{\mathrm{I}}^{1/2}\widetilde{\mathbf{f}}, \\
\mathbf{h}&=\mathbf{R}_{\mathrm{B}}^{1/2}\widetilde{\mathbf{h}},
\end{align}
where $\widetilde{\mathbf{G}}\in\mathbb{C}^{M\times L}$, $\widetilde{\mathbf{f}}\in\mathbb{C}^{L\times 1}$ and $\widetilde{\mathbf{h}}\in\mathbb{C}^{M\times 1}$ consist of uncorrelated elements $\mathcal{CN}(0, \beta_{\mathrm{G}})$, $\mathcal{CN}(0, \beta_{\mathrm{f}})$  and $\mathcal{CN}(0, \beta_{\mathrm{h}})$, respectively, where $\beta_{\mathrm{G}}$, $\beta_{\mathrm{f}}$ and $\beta_{\mathrm{h}}$ are the path loss of the BS-IRS, IRS-UE and BS-UE channels, respectively.

\subsection{IRS Assisted Channel Probing}
 The channel probing is composed of bidirectional measurements, i.e., uplink and downlink probings.
\subsubsection{Uplink Channel Probing}
During the uplink phase, Bob transmits a pilot signal $s_{u}\in \mathbb{C}^{1\times 1}$  to Alice. The received signal at Alice is given by
\begin{align} 
\mathbf{y}_a = \sqrt{P_{b}}\left(\mathbf{h} + \mathbf{G}\mathbf{\Theta}\mathbf{f}\right)s_{u} +\mathbf{n}_a, 
\end{align}
where $\mathbf{y}_a\in \mathbb{C}^{M\times 1}$, $P_{b}$ is the transmit power of Bob, $\mathbf{\Theta}=\text{diag}\left(e^{j\theta_{1}}, e^{j\theta_{2}}, \dots ,e^{j\theta_{L}}\right)$ denotes the IRS
  reflecting-coefficient matrix with $\theta_{l}\in [0, 2\pi]$ for the $l$-th element and $\mathbf{n}_a\in \mathbb{C}^{M\times 1}$ is the complex Gaussian noise vector at Alice. Then Alice applies the precoding matrix $\mathbf{P}\in \mathbb{C}^{M\times M}$ to $\mathbf{y}_a$ and obtains
\begin{align} 
\widehat{\mathbf{y}}_a =\sqrt{P_{b}}\mathbf{P}^{T} \left(\mathbf{h} + \mathbf{G}\mathbf{\Theta}\mathbf{f}\right)s_{u} + \mathbf{P}^{T}\mathbf{n}_a.
\end{align}
After the least square (LS) channel estimation, the uplink channel is estimated as  
\begin{align} \label{y_{a}}
\widetilde{\mathbf{y}}_a =\sqrt{P_{b}}\mathbf{P}^{T} \left(\mathbf{h} + \mathbf{G}\mathbf{\Theta}\mathbf{f}\right) + \mathbf{P}^{T}\mathbf{n}_a s_{u}^{*}.
\end{align}
\subsubsection{Downlink Channel Probing}
During the downlink phase, Alice transmits a pilot signal of length $M$, $\mathbf{S}_{d}\in \mathbb{C}^{M\times M}$, with $\mathbf{S}_{d}^H\mathbf{S}_{d} = \mathbf{I}_{M}$ to Bob. The received signal at Bob is given by
\begin{align} 
\mathbf{y}_b = \left(\mathbf{h} + \mathbf{G}\mathbf{\Theta}\mathbf{f}\right)^{T}\mathbf{P}\mathbf{S}_{d}^{H} +\mathbf{n}_b, 
\end{align}
where $\mathbf{y}_b\in \mathbb{C}^{1\times M}$,  $\mathbf{n}_b\in \mathbb{C}^{1\times M}$ is the complex Gaussian noise vector at Bob. After LS channel estimation, the downlink channel is estimated as  
\begin{align} 
\widetilde{\mathbf{y}}_b &= \left(\mathbf{h} + \mathbf{G}\mathbf{\Theta}\mathbf{f}\right)^{T}\mathbf{P}\mathbf{S}_{d}^{H} \mathbf{S}_{d}\left(\mathbf{S}_{d}^{H} \mathbf{S}_{d}\right)^{-1} \!+\! \mathbf{n}_b\mathbf{S}_{d}\left(\mathbf{S}_{d}^{H} \mathbf{S}_{d}\right)^{-1} \nonumber \\
&= \left(\mathbf{h} + \mathbf{G}\mathbf{\Theta}\mathbf{f}\right)^{T}\mathbf{P} + \mathbf{n}_b\mathbf{S}_{d}.
\end{align}
Then Bob transposes $\widetilde{\mathbf{y}}_b$ and obtains 
\begin{align}  \label{y_{b}}
\widetilde{\mathbf{y}}_b^{T} = \mathbf{P}^{T}\left(\mathbf{h} + \mathbf{G}\mathbf{\Theta}\mathbf{f}\right) + \mathbf{S}_{d}^{T}\mathbf{n}_b^{T}.
\end{align}

\section{Secret Key Rate}
In this paper, we assume that the passive Eve is far away from both Alice and Bob and the eavesdropping channels are independent of the channels between Alice and Bob~\cite{Access1}. The scenario with correlated eavesdropping channels will be investigated in our future work. Under this assumption, the SKR can be calculated by the mutual information of legitimate nodes’ channel estimations as follows:
\begin{align} \label{SKR}
I\left(\widetilde{\mathbf{y}}_a;\widetilde{\mathbf{y}}_b^{T}\right) & =H\left(\widetilde{\mathbf{y}}_a\right)+H\left(\widetilde{\mathbf{y}}_b^{T}\right)-H\left(\widetilde{\mathbf{y}}_a,\widetilde{\mathbf{y}}_b^{T}\right),
\end{align}
where $I(X;Y)$ denotes the mutual information between random variables $X$ and $Y$, and $H(\cdot)$ denotes differential entropy. 

To facilitate further derivations, we introduce cascaded channel $\mathbf{h}_{c}\in \mathbb{C}^{M(L+1)\times 1}$, which is given by
\begin{align}
   \mathbf{h}_{c}=\text{vec}([\mathbf{h}~\mathbf{G}\text{diag}(\mathbf{f})]).
\end{align}
Then the combined channel can be expressed in a compact form:
\begin{align} \label{compact}
   \mathbf{h} + \mathbf{G}\mathbf{\Theta}\mathbf{f}\overset{(a)}{=}\begin{bmatrix}\mathbf{h}~ \mathbf{G}\text{diag}(\mathbf{f})\end{bmatrix}
       \begin{bmatrix}
         1 \\
        \bm{\theta}
       \end{bmatrix}
       \overset{(b)}{=}(\widetilde{\bm{\theta}}^{T}\otimes\mathbf{I}_{M})\mathbf{h}_{c},
\end{align}
where $\bm{\theta}=\left[[\mathbf{\Theta}]_{1,1}, [\mathbf{\Theta}]_{2,2},\cdots, [\mathbf{\Theta}]_{L,L}\right]^T$, $\widetilde{\bm{\theta}}=[1,\bm{\theta}^T]^T$, $(a)$ holds because $\text{diag}(\mathbf{\bm{\theta}})\mathbf{f}=\text{diag}(\mathbf{f})\bm{\theta}$ and $(b)$ holds because $\text{vec}(\mathbf{X}\mathbf{Y}\mathbf{Z})=(\mathbf{Z}^T\otimes \mathbf{X})\text{vec}(\mathbf{Y})$. Plugging (\ref{compact}) into (\ref{y_{a}}) and (\ref{y_{b}}), we have 
\begin{align} 
\widetilde{\mathbf{y}}_a &= \sqrt{P_{b}}\left(\widetilde{\bm{\theta}}^{T}\otimes\mathbf{P}^{T}\right)\mathbf{h}_{c} + \mathbf{P}^{T}\mathbf{n}_a s_{u}^{*} \nonumber \\
&= \sqrt{P_{b}}\left(\widetilde{\bm{\theta}}\otimes\mathbf{P}\right)^T\mathbf{h}_{c} + \mathbf{P}^{T}\mathbf{n}_a s_{u}^{*}, \\
\widetilde{\mathbf{y}}_b^{T} &= \left(\widetilde{\bm{\theta}}\otimes\mathbf{P}\right)^T\mathbf{h}_{c} + \mathbf{S}_{d}^{T}\mathbf{n}_b^{T}.
\end{align}
\begin{theorem}
The SKR of the IRS assisted multi-antenna system is given by
\begin{align}
I\left(\widetilde{\mathbf{y}}_a;\widetilde{\mathbf{y}}_b^{T}\right) = \log_2 \left(\frac{\left|P_{b}\mathbf{R}_{\mathrm{Z}} + \delta^{2}\mathbf{P}^{T}\mathbf{P}^{*}\right|\left|\mathbf{R}_{\mathrm{Z}} + \delta^{2}\mathbf{I}_{M}\right|}{\left| P_{b}\delta^{2}\mathbf{R}_{\mathrm{Z}} + \delta^{2}\mathbf{P}^{T}\mathbf{P}^{*}\left(\mathbf{R}_{\mathrm{Z}} + \delta^{2}\mathbf{I}_{M}\right)\right|}\right),
\end{align}
where $\delta^{2}$ is the Gaussian noise power and $\mathbf{R}_{\mathrm{Z}}=\left(\widetilde{\bm{\theta}}\otimes\mathbf{P}\right)^T\mathbf{R}_c\left(\widetilde{\bm{\theta}}\otimes\mathbf{P}\right)^*$ with 
\begin{align}
\mathbf{R}_{c}= \begin{bmatrix} \beta_{\mathrm{h}}\mathbf{R}_{\mathrm{B}} & \mathbf{0}^{T} \\
    \mathbf{0} & \beta_{\mathrm{G}}\beta_{\mathrm{f}}\mathbf{R}_{\mathrm{I}}\odot \mathbf{R}_{\mathrm{I}}\otimes \mathbf{R}_{\mathrm{B}}
  \end{bmatrix}.
\end{align}
\end{theorem}
\begin{IEEEproof}
See Appendix \ref{AppendixA}.
\end{IEEEproof}

In this paper, we aim to maximize the SKR by jointly optimizing the BS precoding matrix of $\mathbf{P}$ and the IRS reflecting coefficient vector of  $\bm{\theta}$. The channel spatial covariance is assumed to be known by the BS~\cite{TWC1}. Accordingly, the optimization problem is formulated as
\begin{align}
(\mathcal{P}1)~\mathop{\max_{\mathbf{P}, \bm{\theta}}}~&I\left(\widetilde{\mathbf{y}}_a;\widetilde{\mathbf{y}}_b^{T}\right)  \label{P}   \\
      s.t.~          &\text{Tr}\left(\mathbf{P}\mathbf{P}^H\right) = P_{a}M, \tag{\ref{P}{a}} \label{c1a} \\
      &   \left|[\mathbf{\Theta}]_{l,l}\right| = 1, l=1,2,\cdots,L, \label{c1b} \tag{\ref{P}{b}}
\end{align}
where $P_{a}$ is the transmit power  of Alice.

\section{Baseline Solution}\label{sec:baseline}

The original optimization problem is a high-dimensional non-convex problem, which is difficult to solve with conventional optimization methods. For ease of further simplification, we rewrite $I\left(\widetilde{\mathbf{y}}_a;\widetilde{\mathbf{y}}_b^{T}\right)$ as 
\begin{align}
&I\left(\widetilde{\mathbf{y}}_a;\widetilde{\mathbf{y}}_b^{T}\right)=  \nonumber \\
& \log_2 \left(\frac{\left|P_{b}\delta_{h}^{2}\mathbf{P}^{T}\mathbf{R}_{\mathrm{B}}\mathbf{P}^{*} \!+\! \delta^{2}\mathbf{P}^{T}\mathbf{P}^{*}\right|\left|\delta_{h}^{2}\mathbf{P}^{T}\mathbf{R}_{\mathrm{B}}\mathbf{P}^{*} \!+\! \delta^{2}\mathbf{I}_{M}\right|}{\left| P_{b}\delta^{2}\delta_{h}^{2}\mathbf{P}^{T}\mathbf{R}_{\mathrm{B}}\mathbf{P}^{*} \!+\! \delta^{2}\mathbf{P}^{T}\mathbf{P}^{*}\left(\delta_{h}^{2}\mathbf{P}^{T}\mathbf{R}_{\mathrm{B}}\mathbf{P}^{*} \!+\! \delta^{2}\mathbf{I}_{M}\right)\right|}\right),
\end{align}
where $\delta_{h}^{2}=\beta_{\mathrm{h}} + \beta_{\mathrm{G}}\beta_{\mathrm{f}}\bm{\theta}^H(\mathbf{R}_{\mathrm{I}}\odot\mathbf{R}_{\mathrm{I}})\bm{\theta}$.
We further rewrite $\mathbf{P}$ as $\mathbf{P}=\sqrt{P_{a}}\mathbf{P}_{\mathrm{e}}$, where $\mathbf{P}_{\mathrm{e}}$ is a normalized matrix. Then $I\left(\widetilde{\mathbf{y}}_a;\widetilde{\mathbf{y}}_b^{T}\right)$ can be derived as
\begin{align} \label{SKR2}
&{I}\left(\widetilde{\mathbf{y}}_a;\widetilde{\mathbf{y}}_b^{T}\right)= \nonumber \\
&\log_2 \left(\frac{\left|P_{b}P_{a}\delta_{h}^{2}\widehat{\mathbf{R}}_{\mathrm{Z}} + \delta^{2}P_{a}\mathbf{P_{\mathrm{e}}}^{T}\mathbf{P}_{\mathrm{e}}^{*}\right|\left|P_{a}\delta_{h}^{2}\widehat{\mathbf{R}}_{\mathrm{Z}} + \delta^{2}\mathbf{I}_{M}\right|}{\left| \delta^{2}P_{b}P_{a}\delta_{h}^{2}\widehat{\mathbf{R}}_{\mathrm{Z}} + \delta^{2}P_{a}\mathbf{P_{\mathrm{e}}}^{T}\mathbf{P}_{\mathrm{e}}^{*}\left(P_{a}\delta_{h}^{2}\widehat{\mathbf{R}}_{\mathrm{Z}} + \delta^{2}\mathbf{I}_{M}\right)\right|}\right),
\end{align}
where $\widehat{\mathbf{R}}_{\mathrm{Z}}=\mathbf{P_{\mathrm{e}}}^{T}\mathbf{R}_{\mathrm{B}}\mathbf{P}_{\mathrm{e}}^{*}$. To facilitate further optimization, we approximate $\mathbf{P_{\mathrm{e}}}^{T}\mathbf{P_{\mathrm{e}}}^{*}$ as $\mathbf{I}_{M}$ in the noise term. Accordingly, ${I}\left(\widetilde{\mathbf{y}}_a;\widetilde{\mathbf{y}}_b^{T}\right)$ is approximated as 
\begin{align} \label{SKR3}
&\widehat{I}\left(\widetilde{\mathbf{y}}_a;\widetilde{\mathbf{y}}_b^{T}\right)= \nonumber \\
&\log_2 \left(\frac{\left|P_{b}P_{a}\delta_{h}^{2}\widehat{\mathbf{R}}_{\mathrm{Z}} + \delta^{2}P_{a}\mathbf{I}_{M}\right|\left|P_{a}\delta_{h}^{2}\widehat{\mathbf{R}}_{\mathrm{Z}} + \delta^{2}\mathbf{I}_{M}\right|}{\left| \delta^{2}P_{b}P_{a}\delta_{h}^{2}\widehat{\mathbf{R}}_{\mathrm{Z}} + \delta^{2}P_{a}\left(P_{a}\delta_{h}^{2}\widehat{\mathbf{R}}_{\mathrm{Z}} + \delta^{2}\mathbf{I}_{M}\right)\right|}\right).
\end{align}
After Cholesky factorization, we have
\begin{align} \label{Cholesky}
\widehat{\mathbf{R}}_{\mathrm{Z}}
&=\mathbf{P}_{\mathrm{e}}^T \mathbf{R}_{\mathrm{B}}^{1/2}\left(\mathbf{R}_{\mathrm{B}}^{1/2}\right)^H \mathbf{P}_{\mathrm{e}}^* \nonumber \\ &=\left(\left(\mathbf{R}_\mathrm{B}^{1/2}\right)^H\mathbf{P}_{\mathrm{e}}^*\right)^H\left(\left(\mathbf{R}_\mathrm{B}^{1/2}\right)^H\mathbf{P}_{\mathrm{e}}^*\right).
\end{align}
 Then we perform the following eigenvalue decomposition:
 \begin{align} \label{SVD}
\left(\mathbf{R}_\mathrm{B}^{1/2}\right)^H\mathbf{P}_{\mathrm{e}}^*=\mathbf{U}\mathbf{\Lambda}\mathbf{U}^H,
\end{align}
where  $\mathbf{\Lambda}=\text{diag}(p_1,p_2,\dots,p_M)$ with the eigenvalues sorted in descending order. 
Substituting (\ref{Cholesky}) and (\ref{SVD}) into (\ref{SKR3}), $\widehat{I}\left(\widetilde{\mathbf{y}}_a;\widetilde{\mathbf{y}}_b^{T}\right)$ can be rewritten as (\ref{SKR2-scalar}), which is shown at the top of the next page.
\begin{lemma} \label{lemma1}
$\widehat{I}\left(\widetilde{\mathbf{y}}_a;\widetilde{\mathbf{y}}_b^{T}\right)$ is maximized when all the IRS reflecting elements have the same phase configuration, i.e., $\theta_{i}=\theta_{j}, \forall i\ne j$.
\end{lemma}
\begin{IEEEproof}
Due to the space limitation, the proof will be given in the extended journal version.
\end{IEEEproof}

 Following Lemma \ref{lemma1}, we set $\theta_{i}=\theta_{j}, \forall i\ne j$ and have $\delta_{h}^{2}=\beta_{\mathrm{h}} + \beta_{\mathrm{G}}\beta_{\mathrm{f}}\sum_{i=1}^{L}\sum_{j=1}^{L}\left[\mathbf{R}_{\mathrm{I}}\odot\mathbf{R}_{\mathrm{I}}\right]_{i,j}$.

\begin{figure*}[t]
\begin{align}
\label{SKR2-scalar}
\hspace{0.11in}
\widehat{I}\left(\widetilde{\mathbf{y}}_a;\widetilde{\mathbf{y}}_b^{T}\right)&= 
 \log_2 \left(\frac{\left|P_{b}P_{a}\delta_{h}^{2}\mathbf{U}\mathbf{\Lambda}^T\mathbf{\Lambda}\mathbf{U}^H + \delta^{2}P_{a}\mathbf{I}_{M}\right|\left|P_{a}\delta_{h}^{2}\mathbf{U}\mathbf{\Lambda}^T\mathbf{\Lambda}\mathbf{U}^H + \delta^{2}\mathbf{I}_{M}\right|}{\left| \delta^{2}P_{b}P_{a}\delta_{h}^{2}\mathbf{U}\mathbf{\Lambda}^T\mathbf{\Lambda}\mathbf{U}^H + \delta^{2}P_{a}\left(P_{a}\delta_{h}^{2}\mathbf{U}\mathbf{\Lambda}^T\mathbf{\Lambda}\mathbf{U}^H + \delta^{2}\mathbf{I}_{M}\right)\right|}\right) \nonumber \\
&= \sum_{i=1}^{M}\log_2 \left(\frac{(P_{b}P_{a}\delta_{h}^{2}p_{i}^{2} + \delta^{2}P_{a})\left(P_{a}\delta_{h}^{2}p_{i}^{2}+ \delta^{2}\right)}{\delta^{2}P_{b}P_{a}\delta_{h}^{2}p_{i}^{2}+ \delta^{2}P_{a}^{2}\delta_{h}^{2}p_{i}^{2} + \delta^{4}P_{a}}\right),
\end{align}
\hrulefill
\end{figure*}

We further perform eigenvalue decomposition on $\mathbf{R}_{\mathrm{B}}$ and obtain 
$\mathbf{R}_{\mathrm{B}}=\mathbf{U}_{\mathrm{B}}\mathbf{\Lambda}_{\mathrm{B}}\mathbf{U}_{\mathrm{B}}^H$, where $\mathbf{\Lambda}_{\mathrm{B}}=\text{diag}\left(p_{\mathrm{B},1},p_{\mathrm{B},2},\dots,p_{\mathrm{B},M}\right)$ with the eigenvalues sorted in descending order. The optimal $\mathbf{\Lambda}$ can be obtained by solving the following optimization problem:
\begin{align} 
(\mathcal{P}2)~\mathop{\max_{p_{i}}}~&\sum_{i=1}^{M}\log_2 \left(\frac{(P_{b}P_{a}\delta_{h}^{2}p_{i}^{2} + \delta^{2}P_{a})\left(P_{a}\delta_{h}^{2}p_{i}^{2}+ \delta^{2}\right)}{\delta^{2}P_{b}P_{a}\delta_{h}^{2}p_{i}^{2}+ \delta^{2}P_{a}^{2}\delta_{h}^{2}p_{i}^{2} + \delta^{4}P_{a}}\right) \label{P2}  \\
      &s.t.~\sum_{i=1}^{M}\frac{p_{i}^{2}}{p_{\mathrm{B},i}}=M, \label{c2} \tag{\ref{P2}{a}}
\end{align}
where (\ref{c2}) comes from the constraint $\text{Tr}\left(\mathbf{P}_{\mathrm{e}}\mathbf{P}_{\mathrm{e}}^H\right) =\text{Tr}(\mathbf{\Lambda}_\mathrm{B}^{-1}\mathbf{\Lambda}^2)= M$. ($\mathcal{P}2$) can be solved by using water-filling algorithm~\cite{TSP1}, the details of which are omitted here due to limited space. Denoting the optimal $\mathbf{\Lambda}$ obtained from ($\mathcal{P}2$) by $\mathbf{\Lambda}^{\mathrm{opt}}$, the corresponding $\mathbf{P}_{\mathrm{e}}$ can be computed from (\ref{SVD}) and expressed as $\mathbf{P}_{\mathrm{e}}^{\mathrm{opt}}= \left(\mathbf{R}_\mathrm{B}^{-1/2}\mathbf{U}\mathbf{\Lambda}^{\mathrm{opt}}\mathbf{U}^H\right)^{*}$. Notice that $\mathbf{P}_{\mathrm{e}}^{\mathrm{opt}}$ is suboptimal to the original optimization problem  ($\mathcal{P}1$) due to the approximation in (\ref{SKR3}).

\section{Proposed Machine Learning Based Method}
The baseline solution is an iterative algorithm with a complex structure. In this section, we propose to directly solve ($\mathcal{P}1$) using DNNs. More specifically, we aim to train a DNN to learn the mapping relationship between the location information of Bob and the optimal $\mathbf{P}$ and $\bm{\theta}$ that maximize the SKR.

\begin{figure}[!t]
\centerline{\includegraphics[width=3.3in]{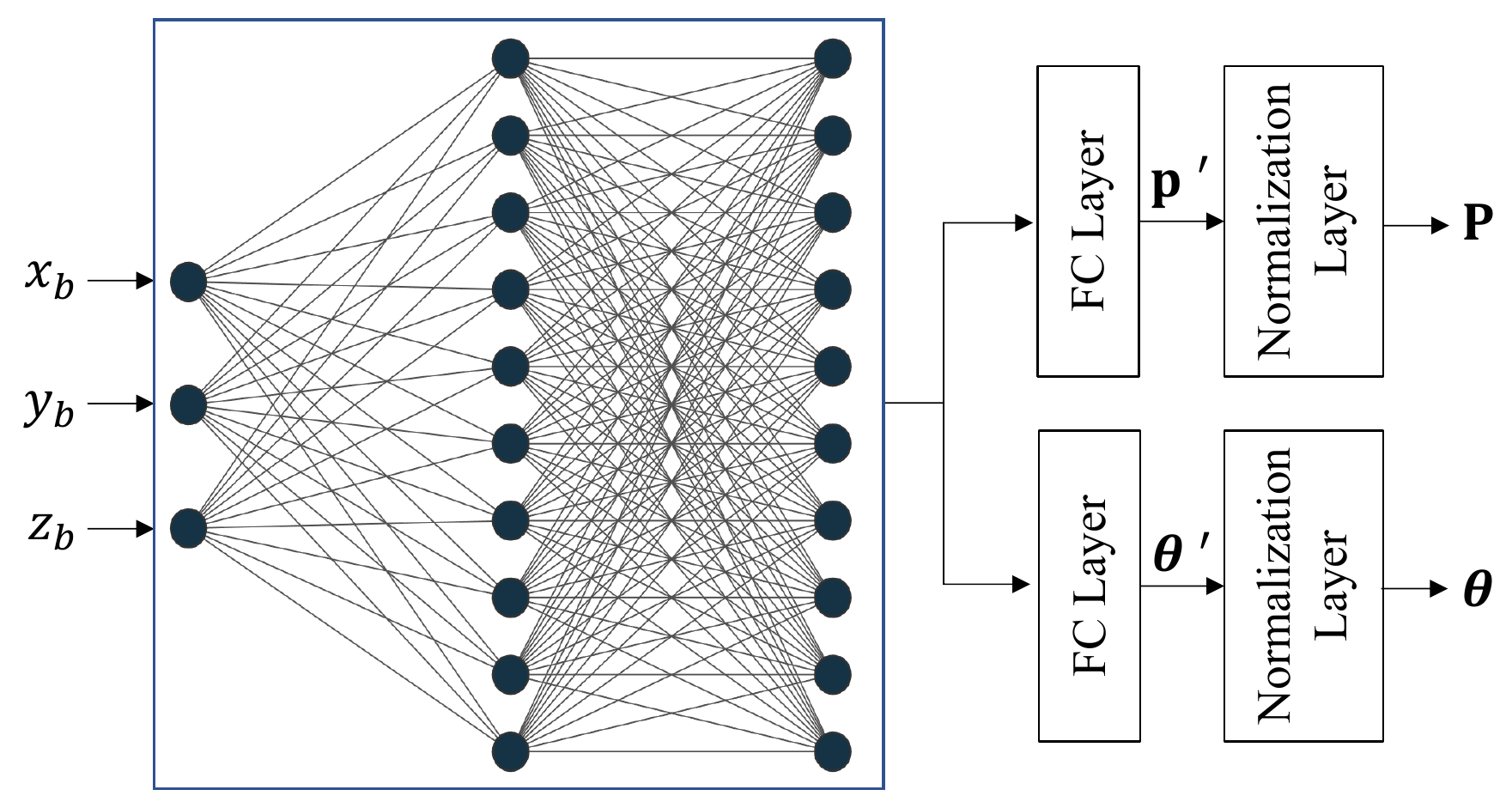}}
\caption{Structure of the proposed PKG-Net.}
\label{fig:network}
\end{figure}

The structure of the proposed PKG-Net is shown in Fig.~\ref{fig:network}. 
In particular, the input is the location information of Bob $\left(x_{b}, y_{b}, z_{b}\right)$, which is a $3\times 1$ tensor. Then two fully connected (FC) layers are adopted as hidden layers for feature extraction with ReLu as the activation function. The output layer consists of two FC layers. Since neural networks only support real-valued outputs, the first $2\times M\times M$ FC layer outputs a real-valued tensor $\mathbf{p}^{'} \in \mathbb{R}^{2M^2\times 1}$ and the second $2\times L$ FC layer outputs a real-valued tensor $\bm{\theta}^{'}\in \mathbb{R}^{2L\times 1}$. To obtain the precoding matrix and satisfy the total power constraint, the first normalization layer converts $\mathbf{p}^{'}$ to a complex-valued matrix $\mathbf{P}$ and performs the following normalization step: 
\begin{align}
\mathbf{P} = \sqrt{MP_{a}}\frac{\mathbf{P}}{\|\mathbf{P}\|_F^2}.
\end{align}
Similarly, to satisfy the unit modulus constraint, the second normalization layer performs 
\begin{align}
\bm{\theta}_{l} = \frac{\bm{\theta}^{'}_{l}}{\sqrt{\left(\bm{\theta}^{'}_{l}\right)^{2} + \left(\bm{\theta}^{'}_{l+L}\right)^{2}}} + j \frac{\bm{\theta}^{'}_{l+L}}{\sqrt{\left(\bm{\theta}^{'}_{l}\right)^{2} + \left(\bm{\theta}^{'}_{l+L}\right)^{2}}}, \forall l.
\end{align}

During the training phase, the PKG-Net learns to update its
parameters in an unsupervised manner. The goal is to maximize the SKR, namely, minimize the following loss function:
\begin{align}
\mathcal{L}_{\mathrm{loss}} = -\frac{1}{K}\sum_{k=1}^{K}I\left(\widetilde{\mathbf{y}}_{a,k};\widetilde{\mathbf{y}}_{b,k}^{T}\right),
\end{align}
where $K$ is the number of training samples, and $\widetilde{\mathbf{y}}_{a,k}$ and $\widetilde{\mathbf{y}}_{b,k}^{T}$ are the channel estimations in the $k$-th training. It is clear that the smaller the loss function, the higher the average SKR. The DNN is updated using the stochastic gradient descent method and implemented by Adam optimizer with a learning rate of 0.001.   The training phase is performed offline, and thus the computational complexity is less a concern.

During the online inference phase, the BS directly designs the precoding matrix and IRS reflecting coefficients according to the output of the trained neural network as soon as it gets the location information of Bob. 

\section{Simulation Results}
In this section, we numerically evaluate the SKR performance of the proposed
PKG-Net in comparison with the benchmark methods.

\subsection{Simulation Setup}
The coordinates of the BS and the IRS in meters are set as (5, -35, 0) and (0, 0 , 0), respectively. The large-scale path loss in dB is computed by 
\begin{align}
	\beta_{i}= \beta_{0} - 10\alpha_{i}\text{log}_{10}\left(d/d_{0}\right),
\end{align}
where $\beta_{0}=-30$ dB is the path loss at the reference distance, $d_{0}=1$~m is the reference distance, $d$ is the transmission distance, and $\alpha_{i}$ is the path-loss exponent.
For the direct link, i.e., $i=\mathrm{h}$, we set $\alpha_{i}=3.67$; for the reflecting links, i.e., $i=\mathrm{G}$ or $\mathrm{f}$, we set $\alpha_{i}=2$. The noise power is set to $\delta^{2}=-90$ dBm. The IRS is assumed to be a uniform square array, i.e., $L_{\mathrm{H}}=L_{\mathrm{V}}$. Unless otherwise specified, the default spatial correlation coefficient at the BS is set to $\eta=0.3$ and the IRS element spacing is half a wavelength, i.e., $\Delta=\frac{\lambda}{2}$. We assume that the uplink and downlink channel probings use the same transmit power, i.e., $P_{a}=P_{b}=P$. 

During the training of the proposed
PKG-Net, the UE location is uniformly distributed in the $xy$-plane with $x\in [5, 15]$ and  $y\in [5, 15]$ to achieve generalization ability to various UE  locations. Each hidden layer of the PKG-Net has 200 neurons.  We train the neural network with 100 epochs.  In each epoch, 1000 random UE locations are used as training samples with a batch size of 100. 

We compare the proposed PKG-Net with the following benchmark methods:
\begin{itemize}
\item Baseline solution: Represent the water-filling algorithm based solution proposed in Section \ref{sec:baseline}.

\item Random configuration: Represent a non-optimized scheme, where both the precoding matrix $\mathbf{P}$ and the IRS reflecting coefficient vector $\bm{\theta}$ are randomly configured.
\end{itemize}

\subsection{Results}
In the following numerical evaluation, the UE location is set as $\left(x_{b}, y_{b}, z_{b}\right)=(10, 10, 0)$ unless otherwise stated. 

\begin{figure}[!t]
\centerline{\includegraphics[width=3.2in]{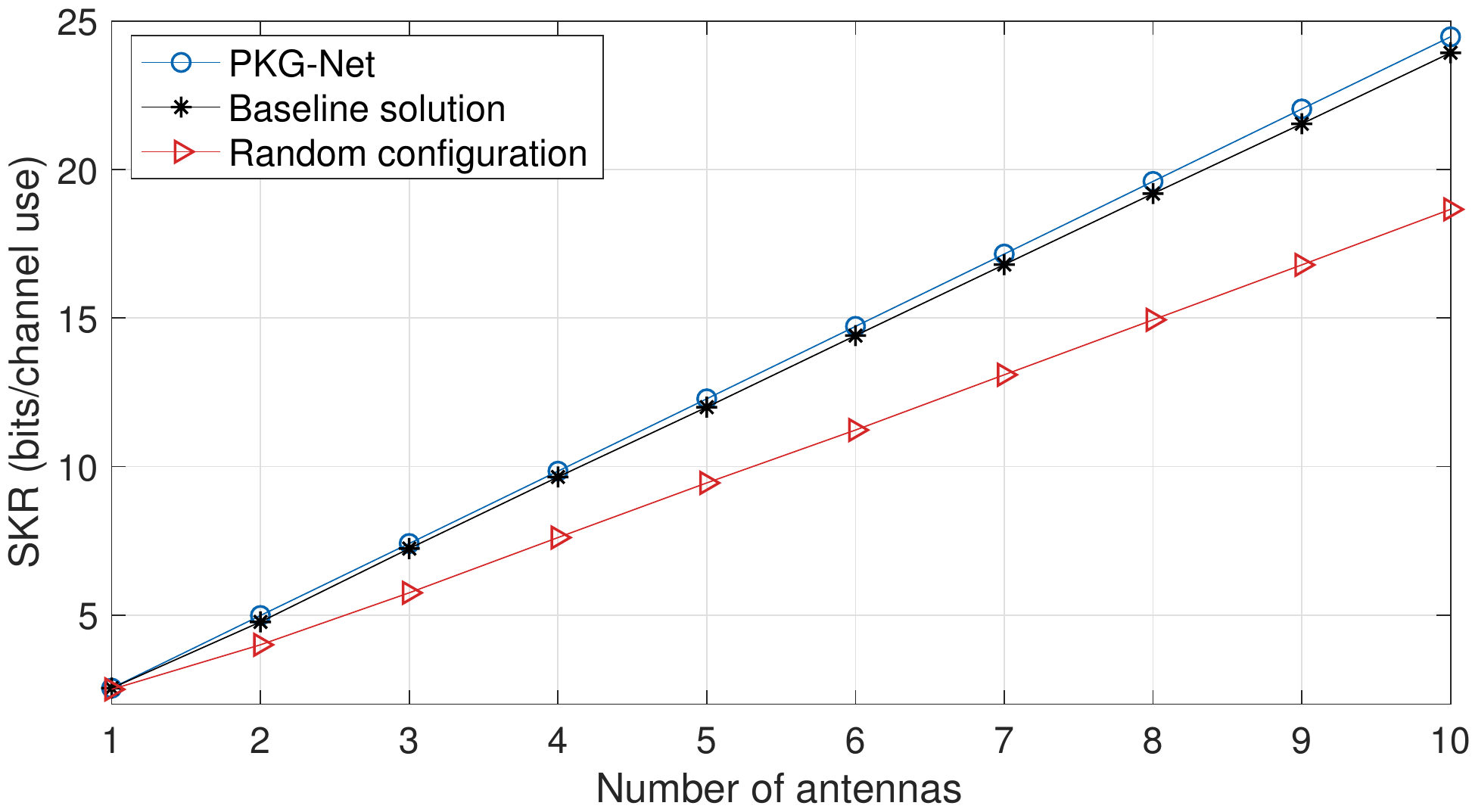}}
\caption{SKR versus $M$ with $P=10$ dBm and $L=36$.}
\label{fig:antenna}
\end{figure}

In Fig. \ref{fig:antenna}, we show the effect of the number of antennas on the SKR. First, it is observed that the SKR  increases linearly with the number of antennas, as more antennas introduce more sub-channels to extract secure keys. Moreover, we can see that regardless of the number of antennas, the proposed PKG-Net achieves a higher SKR than the two benchmark methods, demonstrating the effectiveness of the proposed DNN-based algorithm. The gap in terms of SKR between the proposed PKG-Net and other benchmark methods increases with the number of antennas, indicating that using more antennas requires a more sophisticated design of BS precoding.  

\begin{figure}[!t]
\centerline{\includegraphics[width=3.2in]{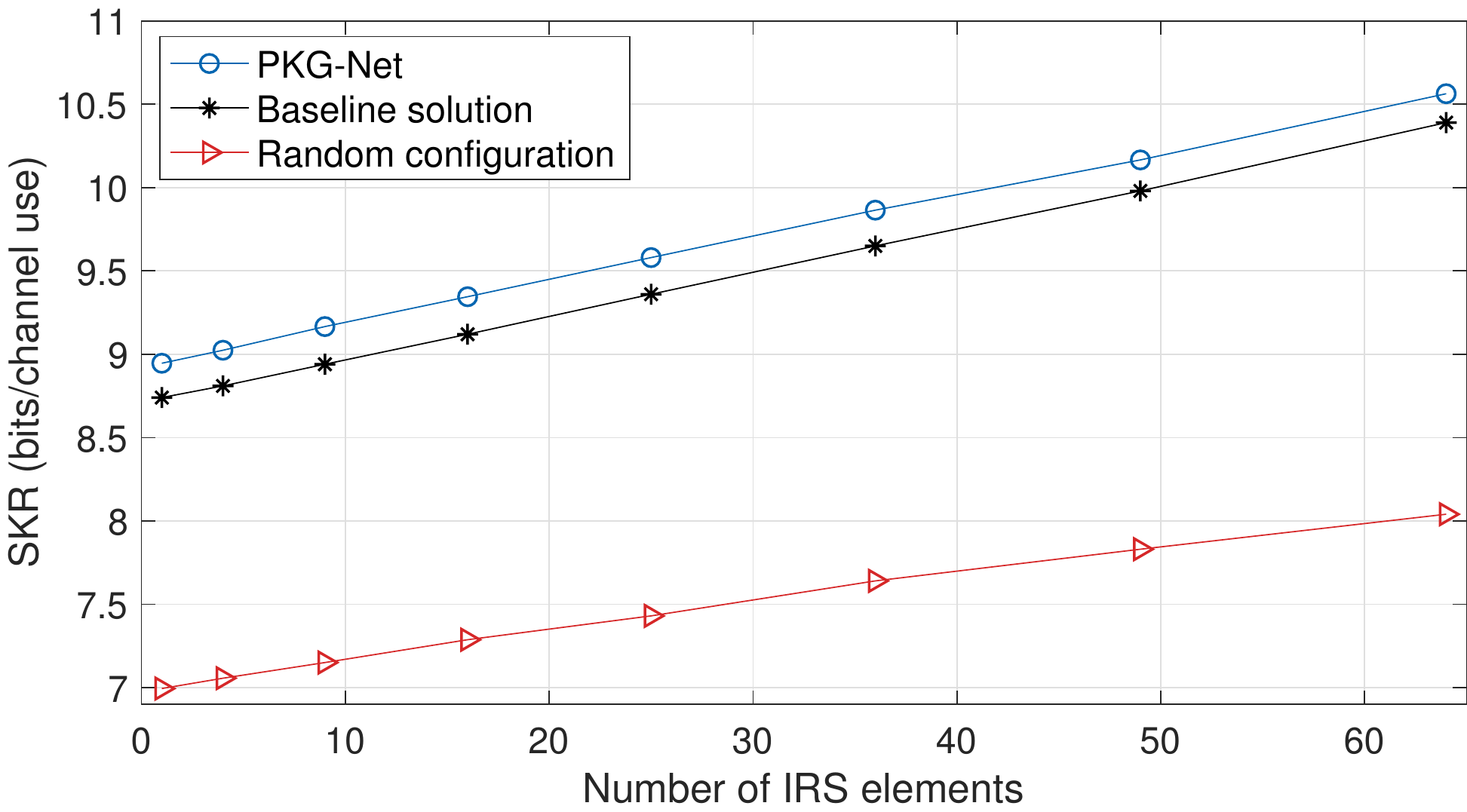}}
\caption{SKR versus $L$ with $P=10$ dBm and $M=4$.}
\label{fig:element}
\end{figure}

Fig. \ref{fig:element} illustrates the effect of the number of IRS elements on the SKR. As can be observed, the SKR monotonically increases with the number of IRS elements, since more IRS elements provide more reflection links that improve the received signals during channel probings. It is noticed that the SKR gain of increasing IRS elements is not as significant as that of increasing BS antennas. This is because in the proposed PKG system, the downlink pilot length is proportional  to the number of antennas. It is expected to achieve a higher SKR when the entire cascaded channel is estimated with a minimum  pilot length of $M(L+1)$, which requires a new design of IRS phase shifting and will be of interest in our future work. 

\begin{figure}[t]
\centerline{\includegraphics[width=3.2in]{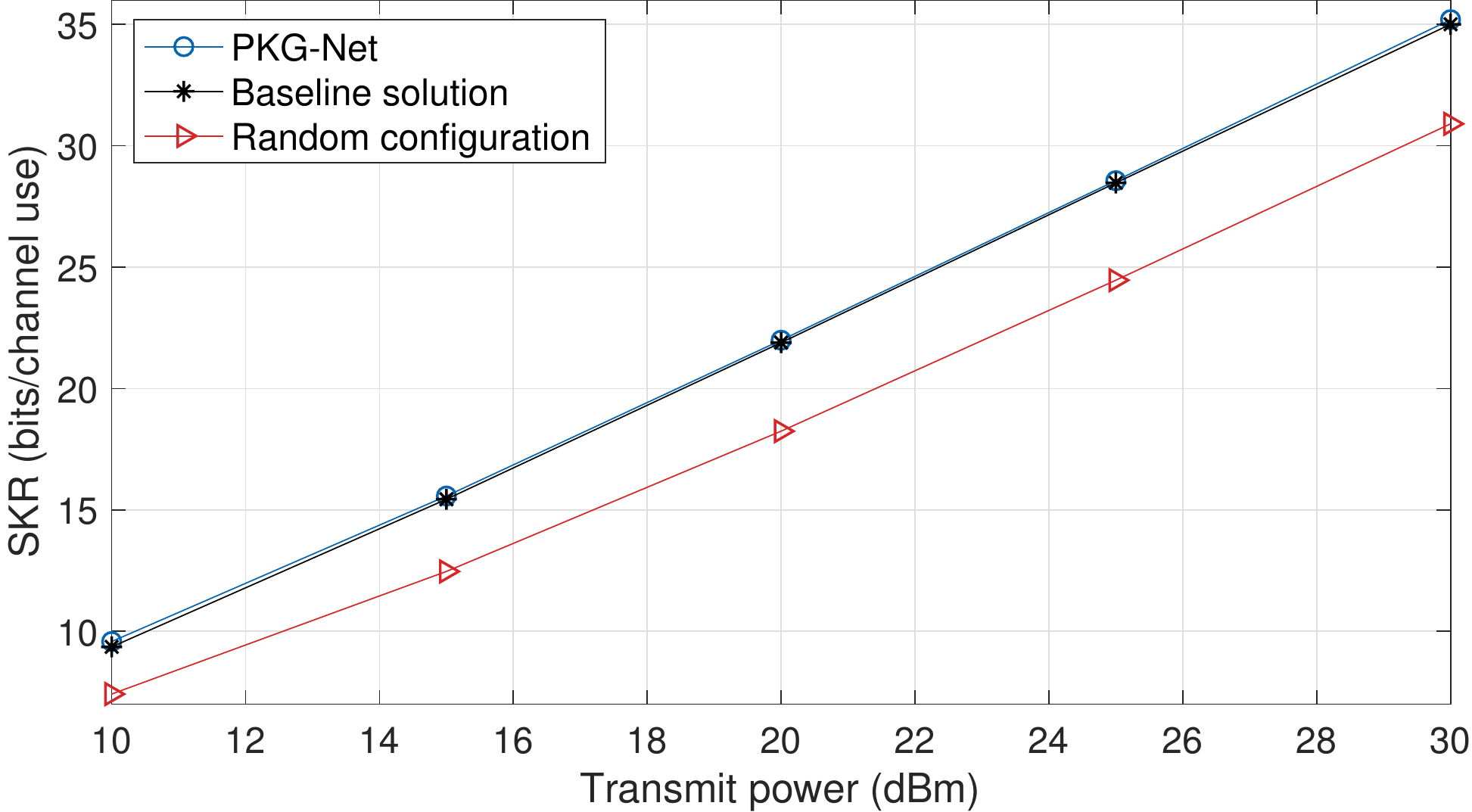}}
\caption{SKR versus $P$ with $M=4$ and $L=25$.}
\label{fig:power}
\end{figure}

In Fig. \ref{fig:power}, the SKR versus the transmit power is displayed. It is shown that the SKR monotonically increases with the transmit power for all the considered optimization methods, which is straightforward due to the fact that a higher transmit power leads to a higher SNR, thereby improving the reciprocity of channel observations at Alice and Bob. 

\begin{figure}[t]
\centerline{\includegraphics[width=3.2in]{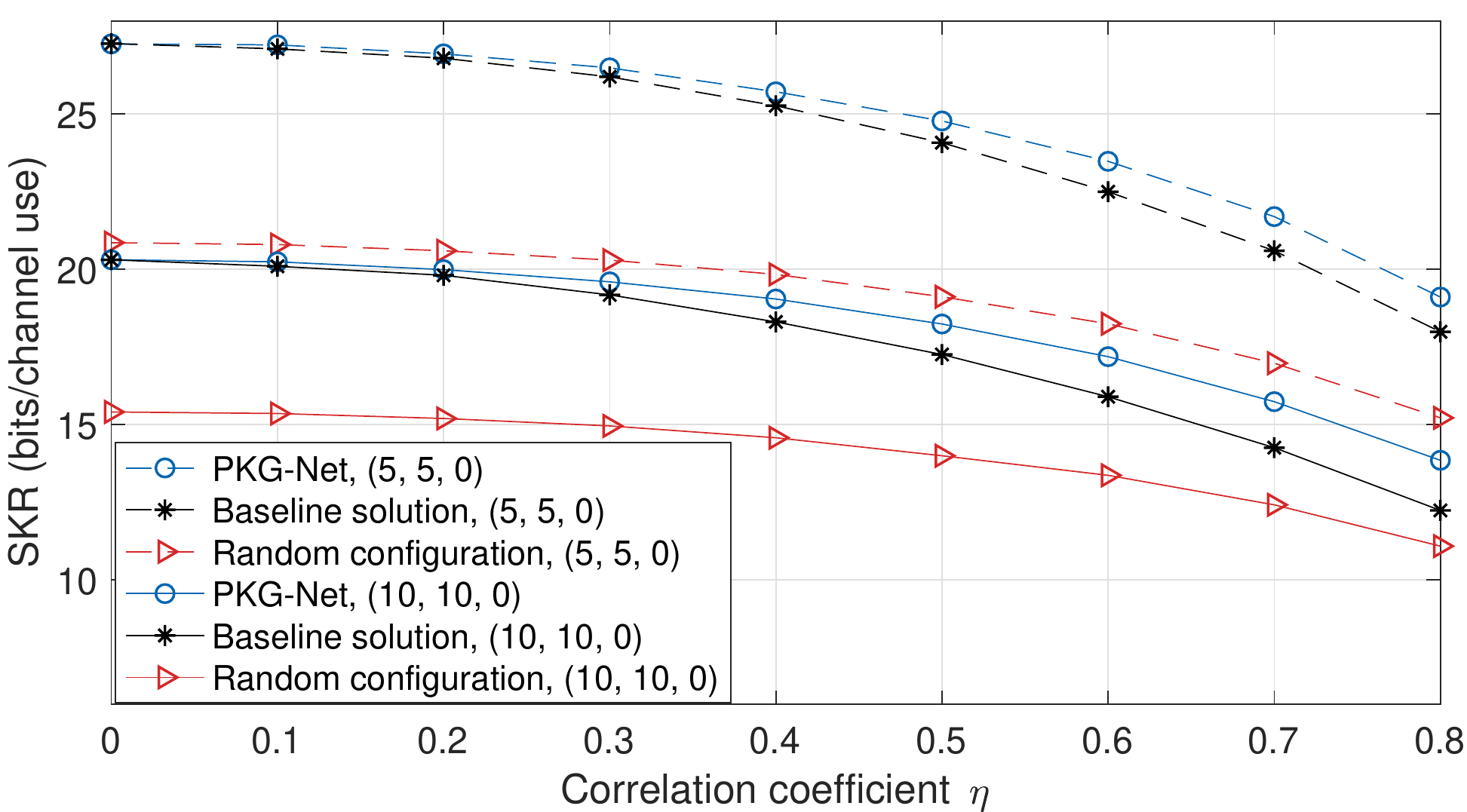}}
\caption{SKR versus $\eta$ with $P=10$ dBm, $M=8$ and $L=36$.}
\label{fig:correlation}
\end{figure}

Fig. \ref{fig:correlation} shows the SKR versus the spatial correlation coefficient at the BS for different UE locations. We can see that the SKR monotonically decreases with the correlation coefficient, which illustrates that the existence of spatial correlation degrades the performance of PKG in terms of SKR. In comparison with the benchmark methods, the proposed PKG-Net can effectively design the BS precoding and IRS phase shifting under different spatial correlation conditions. Interestingly, when the correlation coefficient increases, the SKR gain of the proposed PKG-Net becomes more significant compared to the baseline solution. Furthermore, the proposed PKG-Net shows good generalization to UE locations.

\section{Conclusion}
In this paper, we have studied the PKG in an IRS-assisted multi-antenna system. We have developed a new PKG framework and derived the closed-form expression of SKR.  To maximize the SKR, we proposed a novel DNN-based algorithm, PKG-Net, to jointly configure the precoding matrix at the BS  and the reflecting coefficient vector at the IRS. Simulation results show that under various system parameters, the proposed PKG-Net achieves a higher SKR than the benchmark methods. Additionally, it is observed that the spatial correlation among BS antennas degrades the PKG performance in terms of SKR, and that the proposed PKG-Net can address the spatial correlation more effectively than other benchmark methods. In our future work, we will extend the proposed PKG system to a multi-user scenario and solve the multi-user precoding problem.

\appendices
\section{Proof of Theorem 1} \label{AppendixA}
Following~\cite{TWC1}, (\ref{SKR}) can be further derived as
\begin{align} 
I\left(\widetilde{\mathbf{y}}_a;\widetilde{\mathbf{y}}_b^{T}\right) & =H\left(\widetilde{\mathbf{y}}_a\right)+H\left(\widetilde{\mathbf{y}}_b^{T}\right)-H\left(\widetilde{\mathbf{y}}_a,\widetilde{\mathbf{y}}_b^{T}\right) \nonumber \\
& = \log_2 \left(\frac{|\mathbf{R}_{a}||\mathbf{R}_{b}|}{\left|\bm{\mathcal{R}}_{a,b}\right|}\right),
\end{align}
where 
\begin{align} 
\mathbf{R}_{a}&= \mathbb{E}\left\{\widetilde{\mathbf{y}}_{a} \widetilde{\mathbf{y}}_{a}^H\right\}= P_{b}\left(\widetilde{\bm{\theta}}\otimes\mathbf{P}\right)^T\mathbf{R}_c\left(\widetilde{\bm{\theta}}\otimes\mathbf{P}\right)^*+\delta^{2}\mathbf{P}^{T}\mathbf{P}^{*},
\\
\mathbf{R}_{b}&=\mathbb{E}\left\{\widetilde{\mathbf{y}}_{b}^{T} \widetilde{\mathbf{y}}_{b}^{*}\right\}=\left(\widetilde{\bm{\theta}}\otimes\mathbf{P}\right)^T\mathbf{R}_c\left(\widetilde{\bm{\theta}}\otimes\mathbf{P}\right)^*+\delta^{2}\mathbf{I}_{M},
\end{align}
and
\begin{align} 
\bm{\mathcal{R}}_{a,b}=\begin{bmatrix} \mathbf{R}_{a} &\mathbf{R}_{a,b} \\
    \mathbf{R}_{b,a} &\mathbf{R}_{b}
  \end{bmatrix},
\end{align}
where 
$\mathbf{R}_c = \mathbb{E}\left\{\mathbf{h}_{c}\mathbf{h}_{c}^H\right\}= \begin{bmatrix} \beta_{\mathrm{h}}\mathbf{R}_{\mathrm{B}} & \mathbf{0}^{T} \\
    \mathbf{0} & \beta_{\mathrm{G}}\beta_{\mathrm{f}}\mathbf{R}_{\mathrm{I}}\odot \mathbf{R}_{\mathrm{I}}\otimes \mathbf{R}_{\mathrm{B}}
  \end{bmatrix}$ and 
\begin{align} 
\mathbf{R}_{a, b}=\mathbf{R}_{b, a}=\mathbb{E}\left\{\widetilde{\mathbf{y}}_{a} \widetilde{\mathbf{y}}_{b}^{*}\right\}= \sqrt{P_{b}}\left(\widetilde{\bm{\theta}}\otimes\mathbf{P}\right)^T\mathbf{R}_c\left(\widetilde{\bm{\theta}}\otimes\mathbf{P}\right)^*.
\end{align}
This completes the proof.

\section*{Acknowledgement}
The work of C. Chen and J. Zhang was in part supported by the UK EPSRC under grant ID EP/V027697/1. The work of J.~Zhang, T. Lu and L. Chen was in part supported by National Key Research, Development Program of China under grant ID 2020YFE0200600.



\end{document}